\documentclass[12pt,aps,pre]{revtex4-1}
\usepackage{graphicx}
\usepackage{graphics}
\usepackage{indentfirst}
\usepackage{subfigure}
\usepackage{amsmath, amsthm, amssymb}
\usepackage[T1]{fontenc}
\usepackage[latin9]{inputenc}

\begin{document}

\title{Topological structures in the Husimi flow}

\author{M.~Veronez and M.~A.~M.~de Aguiar}

\affiliation{Instituto de F\'{\i}sica `Gleb Wataghin', Universidade
Estadual de Campinas (UNICAMP)\\ 13083-970, Campinas, Brazil}

\begin{abstract}

We study the topological properties of the quantum phase space current in the Husimi representation, focusing on the dynamical differences, induced by these properties, between the quantum and the classical flows. We show that the zeros of the Husimi function are stagnation points of the current and have a nonzero topological charge. Due to overall charge conservation, new stagnation points with opposite charge appear in pairs in the Husimi current and they have important roles in dynamical processes. As an example we show the topological effect of the zeros in the transmission rate of particle through a potential barrier.

\end{abstract}

\maketitle

\section{Introduction}

\paragraph*{Motivation.}

Since the early days of the quantum theory physicists have looked for ways to represent quantum states as probability distributions in the phase space, a procedure in which information about coordinate and momentum representations is encoded in one single function \cite{weyl27, wigner32, groen46, moyal49, hillery84}. This type of representation, thoroughly employed in classical statistical mechanics, is nowadays widely used to study quantum systems with applications in experimental photonics,
quantum information,
numerical semiclassical methods
and other applied and theoretical fields.

In general, functions $Q$ representing probability densities in the phase space satisfy the continuity equation
\begin{equation}
\frac{\partial Q}{\partial t}+\text{div} \mathbf{J}=\sigma,
\label{eq:ContGeneral}
\end{equation}
where the divergence is taken with respect to the coordinates $x$ and momenta $p$. This equation defines the probability current $\mathbf{J}$ and the generation rates $\sigma$ for $Q$, which can be either positive or negative. In classical mechanics $\sigma = 0$ due to probability conservation, and the equations of motion for the coordinates determine the current \cite{marsden}. For example, in an one-dimensional system
\begin{equation*}
\mathbf{J}=\left( \begin{array}{c} J_{x} \\ J_{p} \end{array} \right) = Q
\left( \begin{array}{c} \dot{x} \\ \dot{p} \end{array} \right),
\end{equation*}
where the dots indicate total derivative with respect to time. Classical solutions of the equation (\ref{eq:ContGeneral}) can be obtained from an initial function $Q$ and are given in terms of the trajectories governed by the equations of motion: $Q \left( x , p ; t \right) = Q \left( x_0 , p_0 ; 0 \right)$; where $x_0$ and $p_0$ are the initial conditions of the trajectory ending at $x$ and $p$ in the time $t$. These classical probability functions have two properties: their marginal distributions are the physical coordinate and momentum distributions and they are positive semi-definite, $Q \geq 0$.

The quantum mechanical analog of this formulation uses phase space functions to represent states but the general solutions have different properties from that of the classical ones. Firstly, it has long been known that any quantum function violates at least one of the two classical properties stated before. For example, the Wigner function \cite{hillery84, ozorio98} has as marginal distributions the wavefunctions of the state in the coordinate and momentum representations but it is not positive, and the Husimi function \cite{perelomov, klauder} is positive, but it does not have the wavefunctions as its marginal distributions. Secondly, quantum solutions are not guided by trajectories because it is impossible to assess both coordinate and momentum for a given test particle, even though the quantum functions satisfy the continuity equation and have well a defined probability current \cite{steuer2013, veronez2013}. Since authentic quantum trajectories do not exist as in the classical dynamics, in order to characterize both dynamical regimes in the phase space on the same grounds we need to attain the analysis on the probability flow itself.

\paragraph*{Quantum phase space flow.}

Continuity equations in quantum mechanics cannot be derived from equations of motion. Instead they are built by casting the von Neumann equation for the dynamics of the states in some representation, in which the density operator of the state is mapped to the function $Q$ in the phase space, where a suitable current $\mathbf{J}$ is obtained. In recent years attention has been drawn to the topological structures the quantum current generates and the relations to its classical counterpart. So far it has been shown that the Wigner current has a strict topological order in the dynamics of its stagnation points \cite{steuer2013}, and in a recent work the same kind of order was explored for the Husimi function \cite{veronez2013}.

The Husimi function is obtained by averaging the density operator in a basis of Gaussian coherent states and it can be interpreted as the probability of measuring the position and momentum of the quantum system within an uncertainty area centered in the basis state. This function inherits the analiticity of the coherent states, which restricts the dynamics of the stagnation points. In a previous work \cite{veronez2013} we presented a detailed demonstration of the construction of the Husimi current for one-dimensional systems, and we showed that the current adds new topological structures to the dynamics when compared with its classical counterpart. We also conjectured that these new structures emerge because every zero of the Husimi function is also a saddle point of the flow. In this paper we prove this conjecture and we explore the effects of this topological structures in a toy model.

Specifically, we consider transmission through a gaussian barrier and show the connection between the classical and quantum transmission coefficients and the position of the zeros relative to the peaks of Husimi function and the classical energy levels. It has already been observed that the emergence of zeros in the phase space is a signature of quantum interference \cite{leboeuf90, cibils92}; our work adds another layer of comprehension of this signature by analysing a dynamical feature of the zeros.

\paragraph*{Structure of the paper.}

In section 2 we present a summary of our previous work \cite{veronez2013}. Section 3 contains the demonstration of the conjecture for the two-dimensional phase space. In section 4 we use the Gaussian barrier as a toy model to analyse the dynamical effect of the topological structures. Section 5 contains the final remarks.

\section{The Husimi flow}

The coherent states $\vert z \rangle$ of the harmonic oscillator of mass $m$ and frequency $\omega$ are defined as the eigenstates of the annihilation operator,
\begin{equation*}
\widehat{a} \vert z \rangle = z \vert z \rangle,
\end{equation*}
where $z$ is a complex number that labels the eigenvalue. The normalized coherent states can be written as
\begin{equation*}
\vert z \rangle = e^{- \bar{z} z / 2} e^{z \widehat{a}^{\dagger}} \vert 0 \rangle,
\end{equation*}
where $\bar{z}$ is the complex conjugate of $z$ and $\vert 0 \rangle$ is the ground state of the harmonic oscillator. Throughout this paper we use a bar to denote the complex conjugate. The annihilation and creation operators, $\widehat{a}$ and $\widehat{a}^{\dagger}$, are given by
\begin{equation*}
\widehat{a} = \frac{\widehat{x}}{\sigma_{x}} + i \frac{\widehat{p}}{\sigma_{p}} , \qquad \widehat{a}^{\dagger} = \frac{\widehat{x}}{\sigma_{x}} - i \frac{\widehat{p}}{\sigma_{p}},
\end{equation*}
where $\sigma_{x} = \sqrt{\hbar / 2 m \omega}$ and $\sigma_{p} = \sqrt{\hbar m \omega / 2}$ are the coordinate and momentum widths of the ground state, respectively \cite{perelomov, klauder}. The coherent states are minimum uncertainty localized Gaussian wavepackets centered at $x = \left \langle z \vert \widehat{x} \vert z \right \rangle = \sigma_{x} \mathfrak{Re} \left( z \right)$ and $p = \left \langle z \vert \widehat{p} \vert z \right \rangle = \sigma_{p} \mathfrak{Im} \left( z \right)$. In terms of $x$ and $p$ we can rewrite $z$ as
\begin{equation}
z = \frac{x}{\sigma_{x}} + i \frac{p}{\sigma_{p}} , \qquad \bar{z} = \frac{x}{\sigma_{x}} - i \frac{p}{\sigma_{p}}.
\label{eq:definitionZxp}
\end{equation}
The change of variables $\left( x , p \right) \mapsto \left( z , i \hbar \bar{z} \right)$ is a canonical transformation and due to the scaling by $\sigma_{x}$ and $\sigma_{p}$ the dimensionless $z$ coordinate have characteristic size of $1 / \sqrt{\hbar}$, which is important when expansions in powers of $\hbar$ are needed.

The identity operator in the coherent state representation is 
\begin{equation*}
\widehat{\mathbf{1}} = \int \text{d}^{2} z \vert z \rangle \langle z \vert,
\end{equation*}
where the $\text{d}^{2} z = \text{d} \bar{z} \text{d} z / 2 \pi i = \text{d} x \text{d} p / 2 \pi \hbar$ is the displacement invariant volume of the phase space. A quantum state represented by the density operator $\widehat{\rho}$ can be mapped to a phase space function by means of its average over coherent states:
\begin{equation}
Q \left( \bar{z} , z \right) = \text{tr} \left( \widehat{\rho} \vert z \rangle \langle z \vert \right).
\label{eq:definitionQ}
\end{equation}
This is the Husimi function of the quantum state, also called its $Q$-symbol. If the state is pure the Husimi function is the squared modulus of the wavefunction and is non-negative. It represents the minimal uncertainty probability of measuring the position and momentum of the state. From the definition of $\vert z\rangle$, the wavefunction in the coherent state representation can be written as
\begin{equation}
\psi \left( \bar{z} , z \right) \equiv \langle z \vert \psi \rangle = e^{-\bar{z} z / 2} \theta \left( \bar{z} \right),
\label{eq:definitionTheta}
\end{equation}
where $\theta \left( \bar{z} \right) = \langle 0 \vert e^{\bar{z} \widehat{a}} \vert \psi \rangle$ is an analytic function of $\bar{z}$ \cite{perelomov, klauder}. Similarly, $\bar{\psi} \left( z , \bar{z} \right) = \langle \psi \vert z \rangle = e^{-\bar{z} z / 2} \bar{\theta} \left( z \right)$. This factorization of the wave function will be important in the next
section.

The dynamics of the density operator is governed by the von Neumann equation,
\begin{equation}
i \hbar \frac{\partial\widehat{\rho}}{\partial t} = \left[ \widehat{H} , \widehat{\rho} \right],
\label{eq:definitionVonN}
\end{equation}
where $\widehat{H}$ is the Hamiltonian operator and $\left[ \cdot , \cdot \right]$ is the commutator. In order to represent the dynamics in the phase space, we define a Hamiltonian function for $\widehat{H}$,
\begin{equation}
H \left( \bar{z} , z \right) = \text{tr} \left( \widehat{H} \vert z \rangle \langle z \vert \right),
\label{eq:definitionH}
\end{equation}
which is an average of the operator over the coherent states. If the Hamiltonian operator can be written as a normal ordered power series in $\widehat{a}$ and $\widehat{a}^{\dagger}$, namely $\widehat{H} = \sum_{m,n} h_{mn} \widehat{a}^{\dagger m} \widehat{a}^{n}$, then its averaged function is $H = \sum_{m,n} h_{mn} \bar{z}^{m} z^{n}$. With this definition, equation (\ref{eq:definitionVonN}) can be rewritten as a differential equation in the phase space for the Husimi function:
\begin{equation*}
i \hbar \frac{\partial Q}{\partial t} = \sum_{m,n} h_{mn} \bar{z}^{m} \left( \frac{\partial}{\partial\bar{z}} + z \right)^{n} Q - \text{c.c.},
\end{equation*}
where $\text{c.c.}$ stands for the complex conjugate of the expression immediately preceding it. Taking the limit $\hbar \rightarrow 0$ the right hand side of this equation reduces to the Poisson bracket, and the classical dynamics is obtained. In this limit the evolution of a phase space function can be further written as a continuity equation:
\begin{equation*}
\frac{\partial Q}{\partial t} + \frac{\partial J_{cl}}{\partial z} + \frac{\partial\bar{J}_{cl}}{\partial\bar{z}} = 0,
\end{equation*}
where $J_{cl} \left( \bar{z} , z \right) = \frac{1}{i \hbar} Q \frac{\partial H}{\partial \bar{z}}$ is the classical probability current. This classical dynamics is based on the existence of trajectories, guided by the equation of motion $\dot{z} = \frac{1}{i \hbar} \frac{\partial H}{\partial \bar{z}}$, that carry the $Q$ function over the phase space. In a similar fashion we can transform the full quantum dynamical equation into a sourceless continuity equation if the Hamiltonian operator is Hermitian:
\begin{equation}
\frac{\partial Q}{\partial t} + \frac{\partial J}{\partial z} + \frac{\partial \bar{J}}{\partial \bar{z}} = 0;
\label{eq:definitionCont}
\end{equation}
where $J\left(\bar{z},z\right)$ is the quantum probability current and is given by
\begin{equation}
J = \frac{1}{i \hbar} \sum_{m,n = 0}^{\infty} \sum_{k = 0}^{\min \left( m , n \right)} \sum_{l = 1}^{m - k} \frac{h_{mn} \left( -1 \right)^{k} m! n!}{k! l! \left( m - k - l \right)! \left( n - k \right)!} \frac{\partial^{l - 1}}{\partial z^{l - 1}} \left(\bar{z}^{m - k - l}z^{n - k} Q \right).
\label{eq:definitionJ1}
\end{equation}
The lowest order $\hbar$ term on the right hand side of (\ref{eq:definitionJ1}) is obtained with $l = 1$ and $k = 0$, and this term retrieves the classical current $J_{cl}$. In this way, the quantum current can be separated into the classical current plus higher order corrections in $\hbar$. If the Hamiltonian is not Hermitian, $h_{mn} \neq \bar{h}_{nm}$, a source term
\begin{equation*}
\sigma = \frac{Q}{i \hbar}e^{-\frac{\partial}{\partial \bar{z}}\frac{\partial}{\partial z}} \left( H - \bar{H} \right)
\end{equation*}
is added to the right hand side of (\ref{eq:definitionCont}). This term accounts for the absence of norm conservation of the Husimi function. In this paper we will only consider Hermitian Hamiltonians.

\paragraph*{Simplification of $J$.}

The expression (\ref{eq:definitionJ1}) can be further simplified by changing the summation indexes in the following way:
\begin{equation*}
\sum_{m = 0}^{\infty} \sum_{n = 0}^{\infty} \sum_{k = 0}^{\min \left( m , n \right)} \sum_{l = 1}^{m - k} \mapsto \sum_{k = 0}^{\infty} \sum_{l = 1}^{\infty} \sum_{m = k + l + 1}^{\infty} \sum_{n = k}^{\infty}.
\end{equation*}
Expanding the derivatives in $z$ contained in (\ref{eq:definitionJ1}) and gathering conveniently the terms we obtain
\begin{equation}
J = \frac{1}{i \hbar} \sum_{l = 1}^{\infty} \frac{\partial^{l - 1} Q}{\partial z^{l - 1}} \sum_{k = 0}^{\infty} \frac{\left( - 1 \right)^{k}}{\left( k + l \right)!} \frac{\partial^{2k + l} H}{\partial \bar{z}^{k + l} \partial z^{k}},
\label{eq:definitionJ2}
\end{equation}
which is a more compact and manageable expression than the original (\ref{eq:definitionJ1}). Our interest concerns the stagnation points of the flow, those points of the phase space where $J = 0$.

\section{Stagnation Points of the Flow}

In \cite{veronez2013} it was conjectured that the zeros of the Husimi function are also saddle points of $J$. This means that whenever a zero of the function occurs, there are topological differences between the classical and quantum probability flows induced by these new stagnation points, and the phase space dynamics of these regimes are different. In this section we prove that this conjecture is true, and we analyse how the existence of the saddle points can change the dynamical behaviour of a system.The quantum current (\ref{eq:definitionJ2}), like the classical one, has a gauge freedom under the transformation $J \mapsto J + i \frac{\partial}{\partial \bar{z}} \Phi$, for any phase space valued real function $\Phi$. Apart from this gauge, in this work we regard (\ref{eq:definitionJ2}) as the probability current associated to the Husimi function due to its special factorization (\ref{eq:factorizationJ}) shown below.

Let $z_{0}$ be a stagnation point, $J \left( \bar{z}_{0} , z_{0} \right) = 0$. The probability current can be expanded around $z_{0}$ as
\begin{eqnarray*}
J & \approx & \left( z - z_{0} \right) \left. \frac{\partial J}{\partial z} \right|_{\bar{z}_{0} , z_{0}} + \left( \bar{z} - \bar{z}_{0} \right) \left. \frac{\partial J}{\partial \bar{z}} \right|_{\bar{z}_{0} , z_{0}},\\
\bar{J} & \approx & \left( z - z_{0} \right) \left. \frac{\partial \bar{J}}{\partial z} \right|_{\bar{z}_{0} , z_{0}} + \left( \bar{z} - \bar{z}_{0} \right) \left. \frac{\partial \bar{J}}{\partial \bar{z}} \right|_{\bar{z}_{0} , z_{0}}.
\end{eqnarray*}
The topological behaviour of the current about this point is determined by the eigenvalues $\lambda$ of the matrix $\mathcal{G}$ of the linear coefficients of the expansion. This matrix is also the vector gradient of the current, and is given by
\begin{equation*}
\mathcal{G} \left. \left[ \bar{J} , J \right] \right|_{\bar{z}_{0} , z_{0}} = \left. \left(
\begin{array}{cc}
\frac{\partial J}{\partial z} & \frac{\partial J}{\partial \bar{z}}\\
\frac{\partial \bar{J}}{\partial z} & \frac{\partial \bar{J}}{\partial \bar{z}}
\end{array}
\right) \right|_{\bar{z}_{0} , z_{0}}.
\end{equation*}

As an illustration consider the classical case, where the current is given by $J_{cl} = \frac{1}{i \hbar} Q \frac{\partial H}{\partial \bar{z}}$, with $Q = \left| \psi \left( \bar{z} , z \right) \right|^{2}$. There are two sets of stagnation points, given by $Q = 0$ and by $\frac{\partial H}{\partial \bar{z}} = 0$. The eigenvalues of $\mathcal{G}$ at these points are
\begin{equation}
\begin{cases}
\lambda_{cl} = 0, & \text{if } Q = 0,\\
\lambda_{cl} = \pm \frac{Q}{\hbar} \sqrt{\det \mathcal{K}}, & \text{if } \frac{\partial H}{\partial \bar{z}} = 0;
\end{cases}
\label{eq:lambdaClassical}
\end{equation}
where $\mathcal{K}$ is the Hessian matrix of the Hamiltonian function evaluated at the stagnation point:
\begin{equation*}
\mathcal{K} = \left. \left(
\begin{array}{cc}
\frac{\partial ^{2} H}{\partial z ^{2}} & \frac{\partial ^{2} H}{\partial \bar{z} \partial z}\\
\frac{\partial ^{2} H}{\partial \bar{z} \partial z} & \frac{\partial ^{2} H}{\partial \bar{z} ^{2}}
\end{array}
\right) \right|_{\bar{z}_{0} , z_{0}}.
\end{equation*}
When $Q = 0$ the structure of the stagnation point cannot be inferred. When $\frac{\partial H}{\partial \bar{z}} = 0$ the eigenvalues are either real numbers with opposite sign if the Hessian determinant is positive, or pure imaginary conjugate numbers if the determinant is negative. In the former case the point is a saddle of the flow, with an attractive and a repulsive direction; in the latter, it is a vortex \cite{marsden}.

For each stagnation point $z_{0}$ the topological characterization of the probability current can be done by the index $I$ of the flow, which counts the number of times the current rotates completely while we move clockwise around the point. Taking a loop $L$ around $z_{0}$ containing no other stagnation point, $I$ is calculated as the clockwise integral of the angle $\phi$ between the flow and some fixed reference axis in $L$:
\begin{equation*}
I \left( z_{0} \right) = \frac{1}{2 \pi} \oint_{L} \text{d} \phi.
\end{equation*}
One counterclockwise rotation of $J$ adds $-1$, whereas one clockwise rotation adds $+1$. In general, for saddle points $I = -1$ and for vortices $I = +1$. In this way the index define a topological charge $\pm 1$ for each point \cite{marsden, guillemin}.

When we analyse the quantum flow, the behaviour of stagnation points is different from the classical flow. For pure states, the Husimi function can be factored as
\begin{equation}
Q \left( \bar{z} , z \right) = e^{-\bar{z} z} \theta \left( \bar{z} \right) \bar{\theta} \left( z \right),
\label{eq:factorizationQ}
\end{equation}
where $\theta$ is given in (\ref{eq:definitionTheta}). As the quantum current (\ref{eq:definitionJ2}) depends only on the derivatives of $Q$ with respect to $z$, it can be written as
\begin{eqnarray}
J & = & \theta \left( \bar{z} \right) \frac{1}{i \hbar} \sum_{k \geq 0 , l \geq 1} \frac{\left( -1 \right)^{k}}{\left( k + l \right)!} \frac{\partial^{2k + l} H}{\partial \bar{z}^{k + l} \partial z^{k}} \frac{\partial^{l - 1} \left( e^{-\bar{z} z} \bar{\theta} \left( z \right) \right)}{\partial z^{l - 1}} \nonumber \\
 & = & \theta \left( \bar{z} \right) f \left( \bar{z} , z \right).
\label{eq:factorizationJ}
\end{eqnarray}
Here there are two possibilities that produce $J = 0$: $\theta = 0$ or $f = 0$. The points that satisfy the condition $\theta = 0$, which are the zeros of the Husimi function, will be named the \emph{trivial stagnation points}, while those that satisfy $f = 0$, given by an intricate relation between the phase space functions $Q$ and $H$, will be called the \emph{non-trivial stagnation points}.

The eigenvalues of the vector gradient for both classes of stagnation points can be calculated using the factorization (\ref{eq:factorizationJ}):
\begin{equation*}
\mathcal{G} \equiv \left(
\begin{array}{cc}
\theta \frac{\partial f}{\partial z} & \frac{\partial \left(\theta f \right)}{\partial \bar{z}}\\
\frac{\partial \left( \bar{\theta} \bar{f} \right)}{\partial z} & \bar{\theta} \frac{\partial \bar{f}}{\partial \bar{z}}
\end{array}
\right),
\end{equation*}
and are given by the roots of the secular equation
\begin{equation}
\lambda^{2} - \lambda \left( \theta \frac{\partial f}{\partial z} + \bar{\theta} \frac{\partial \bar{f}}{\partial \bar{z}} \right) + \theta \bar{\theta}\frac{\partial f}{\partial z} \frac{\partial \bar{f}}{\partial \bar{z}} - \left( \theta \frac{\partial f}{\partial \bar{z}} + \frac{\partial \theta}{\partial \bar{z}} f \right) \left( \bar{\theta} \frac{\partial \bar{f}}{\partial z} + \frac{\partial \bar{\theta}}{\partial z} \bar{f} \right) = 0.
\label{eq:eqLambda}
\end{equation}
The solutions of (\ref{eq:eqLambda}) are different for each case considered before and are given by
\begin{equation}
\begin{cases}
\lambda^{\pm} = \pm \left| \frac{\partial \theta}{\partial \bar{z}} f \right| , & \text{if } \theta = 0;\\
\lambda^{\pm} = \frac{1}{2} \left( \text{tr} \mathcal{G} \pm \sqrt{\text{tr}^{2} \mathcal{G} - 4 \det \mathcal{G}} \right) , & \text{if } f = 0.
\end{cases}
\label{eq:lambdaQuantum}
\end{equation}
Therefore, the trivial stagnation points are saddles of the flow, and their indices are equal to $-1$. This is the proof of the previous conjecture. There are two possibilities to the eigenvalues of the non-trivial stagnation points. First, when the term under the square root is positive, the eigenvalues are both negative (positive) numbers, and the stagnation point is an attractive (repulsive) node. Second, when the term inside the root is negative, the eigenvalues are a pair of complex conjugate numbers, and in this case the stagnation point is an attractive (repulsive) spiral if their real parts are negative (positive). For both possibilities, the real parts of the $\lambda^{\pm}$'s add to $- \frac{\partial Q}{\partial t}$ and the index of the point is equal to $+1$.

During the time evolution of the quantum state, the movement of the Husimi function is accompanied by the movement of its zeros. In view of the Poincar\'{e}-Hopf theorem the total index of the flow must be conserved during the dynamics, thus the emergency of a saddle point must always be accompanied by the emergency of a non-trivial stagnation point; for this reason stagnation points exist \emph{in pairs}.

In \cite{veronez2013} it has been observed that the stagnation points in the pair move closely to each other in the phase space and form a structure similar to the one depicted in Figure 1. Since this structure is similar to a dipole with opposite charges, in this work we name it \emph{topological dipole}. In the next section we analyse a toy model where the presence of the topological dipoles works as a signature of differences between two regimes of transmission across a potential barrier.

\begin{figure}[!ht]
\begin{centering}
\includegraphics[scale=0.75]{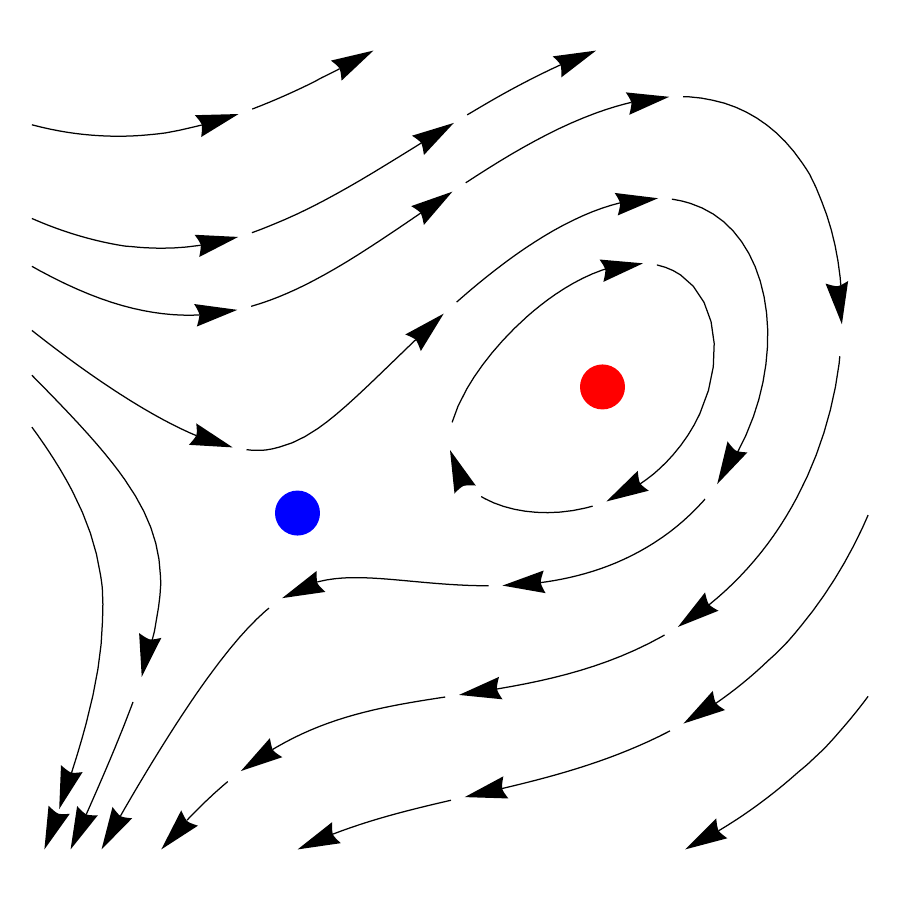}
\par\end{centering}
\caption{{\small Sketch of one topological dipole, comprised by a saddle point (left, blue spot) and a spiral (right, red spot). The total index around both points sums to zero. The vortex can also have an attractive or repulsive character.}}
\label{fig01}
\end{figure}

\section{Tunneling in the Gaussian Barrier}

We consider a particle of mass $m$ scattering off a one-dimensional Gaussian barrier with amplitude $V_{0}$ and width $1/\sqrt{2k}$. We are interested in the comparison between the classical and quantum transmission rates through the barrier, $T_{C}$ and $T_{Q}$, respectively.

The classical Hamiltonian is
\begin{equation}
H_{cl} = \frac{p^{2}}{2m} + V_{0} \exp \left( - k x^{2} \right),
\label{eq:Hclassical}
\end{equation}
and the quantum Hamiltonian for the model is given by
\begin{equation}
\widehat{H} = \frac{\widehat{p}^{2}}{2m} + V_{0} \exp \left( - k \widehat{x}^{2} \right),
\label{eq:Hquantum}
\end{equation}
In general, the classical Hamiltonian is different of the averaged Hamiltonian function defined in (\ref{eq:definitionH}), and the latter contains terms of higher order in $\hbar$. For the operator above, the averaged Hamiltonian (\ref{eq:definitionH}) is given by
\begin{equation}
H = \frac{p^{2}}{2m} + \frac{\hbar \omega}{4} + \alpha V_{0} \exp \left( - \alpha k x^{2} \right),
\label{eq:Haverage}
\end{equation}
where $x$ and $p$ are given by expressions (\ref{eq:definitionZxp}) and $\alpha = \left( 1 + 2k \sigma_{x}^{2} \right)^{-1 / 2}$ is a smoothing factor. The classical Hamiltonian is recovered from the averaged one when $\hbar \rightarrow 0$. The initial state was chosen as a coherent state centered at $x_{0}$ and $p_{0}$ with position and momentum widths $\sigma_{x}$ and $\sigma_{p}$, respectively. The parameters were set to $m = \omega = 1$ and $\hbar = 1 / 100$, which implies $\sigma_{x} = \sigma_{p} = 1/ \left( 10 \sqrt{2} \right)$. We also fixed $V_{0} = 2$, $k = 3$ and $x_{0} = -4$. We also define $p_{C}=\sqrt{2m V_0}=2$, corresponding to a classical kinetic energy equal to the barrier top.  For further details on the numerical calculations we refer the reader to the Appendix.

In order to quantify the relative classical-to-quantum transmission $T$ and reflection $R=1-T$ rates we define
\begin{eqnarray*}
D_{T} & = & \frac{T_{C} - T_{Q}}{T_{C}},\\
D_{R} & = & \frac{R_{C} - R_{Q}}{R_{C}};
\end{eqnarray*}
If $D_{T} > 0$ and $D_{R} < 0$ ($D_{T} < 0$ and $D_{R} > 0$) the probability of the classical particle crossing the barrier is higher (smaller) than the tunneling probability of the quantum particle. There are two different regimes of transmission according to the average initial momentum $p_{0}$ of the particle, as can be seen in Figure 2a. For $p_{0} \lessapprox p_{C}$ the classical transmission is greater than the quantum one, while if $p_{0} \gtrapprox p_{C}$ the quantum transmission is greater. We investigated the structure of the quantum flow for both regimes for a state with initial momentum $p_{0} = 1.8$ and $p_{0} = 2.1$. The quantum movement of the Husimi function is shown for reference in Figure 2b for $p_{0} = 1.8$.

\begin{figure}
\begin{centering}
\includegraphics[scale=0.35]{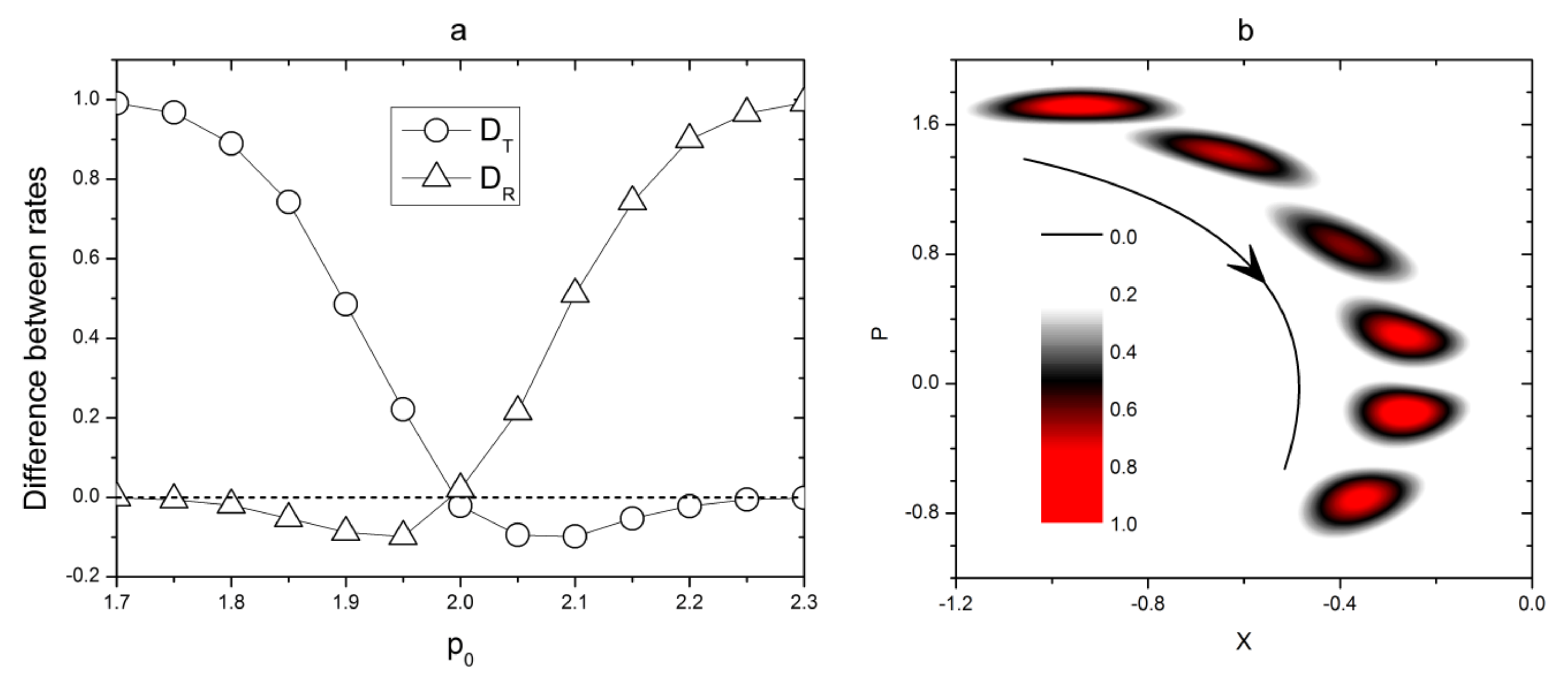}
\par\end{centering}
\caption{{\small (a) Relative transmission and reflection coefficients for different values of $p_{0}$. (b) Husimi function for selected times for $p_{0}=1.8$. Along the arrow, time steps increase by $0.2$ with initial time $1.7$. Amplitude scale relative to the highest value, chosen to be 1.}}
\label{fig02}
\end{figure}

Figure 3 shows the logarithmic plot of the Husimi function, highlighting the position of its zeros, the so called stellar representation \cite{majorana32, leboeuf90, cibils92}. The panels show the Husimi function at different times when the particle is hitting the barrier. In all panels it is possible to see a row of zeros in front of the maximum of the Husimi function (marked by an ellipse with the letters RZ) in a region where classical trajectories cross the potential barrier ($p>0$).

The presence of the zeros causes the quantum flow to be highly distorted with respect to the classical flow. This is seen in Figure 4, which shows the Husimi function superimposed with the quantum current. Three zeros, which are saddles of the current, and their corresponding center companions, are clearly visible (marked by squares, triangles and circles). The flow, that would classically go through the top of the barier to the other side, gets partially blocked by the topological dipoles. This dynamical feature leads to a smaller quantum probability of transmission compared to the classical one.

\begin{figure}
\begin{centering}
\includegraphics[scale=0.25]{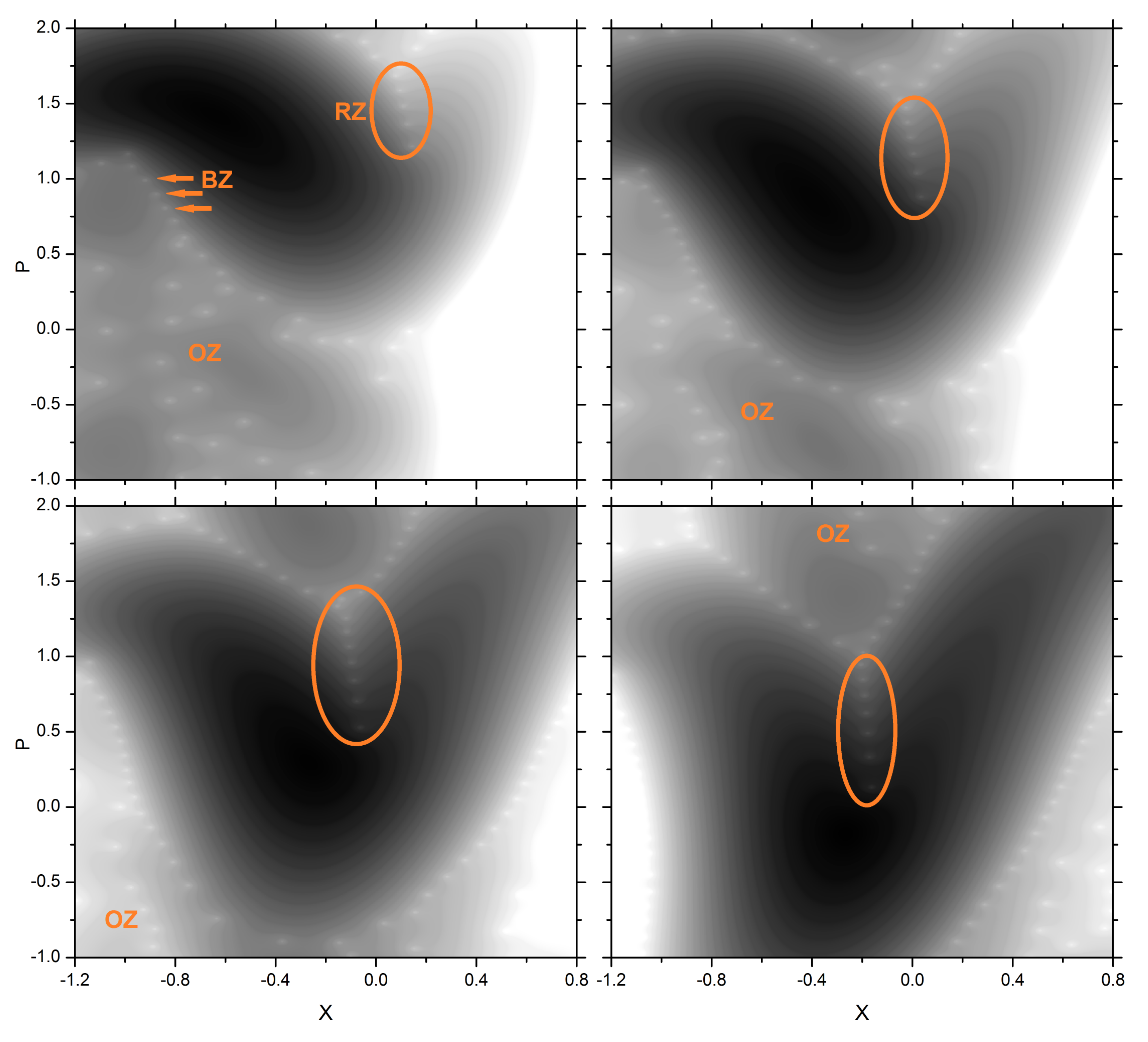}
\par\end{centering}
\caption{{\small Gray scale log-plot of the Husimi function for time instants $T = 1.9 \text{, } 2.1 \text{, } 2.3 \text{ and } 2.5$ (see Figure 2b), organized in the reading direction. Black represents the absolute maximum and white represents zero. A row of zeros (RZ) is seen in front of the Husimi maximal values, which is framed by the border zeros (BZ). The outer zeros (OZ) in the external region are numerical artifacts.}}
\label{fig03}
\end{figure}

In a similar fashion the zeros of the Husimi function and the associated stagnation points of the current help understand the dynamics of the transmission for $p_0 > p_c$, when the quantum transmission is larger than the classical. Figure 4b shows a few classical trajectories superimposed with the Husimi function for $p_{0} = 2.1$. Once again the row of zeros is visible, but this time they are situated in a region near the classical separatrix. Below the separatrix, where the Husimi function is large and the classical trajectories are reflected back, the zeros distort the flow again, allowing  portions of the Husimi function to cross it. Notice that alongside with the last zero of the row, below the separatrix, is the vortex of the topological pair, allowing the flow to circulate around it and move to the other side of the separatrix. This leads to a higher quantum than classical transmission. 

\begin{figure}
\begin{centering}
\includegraphics[scale=0.42]{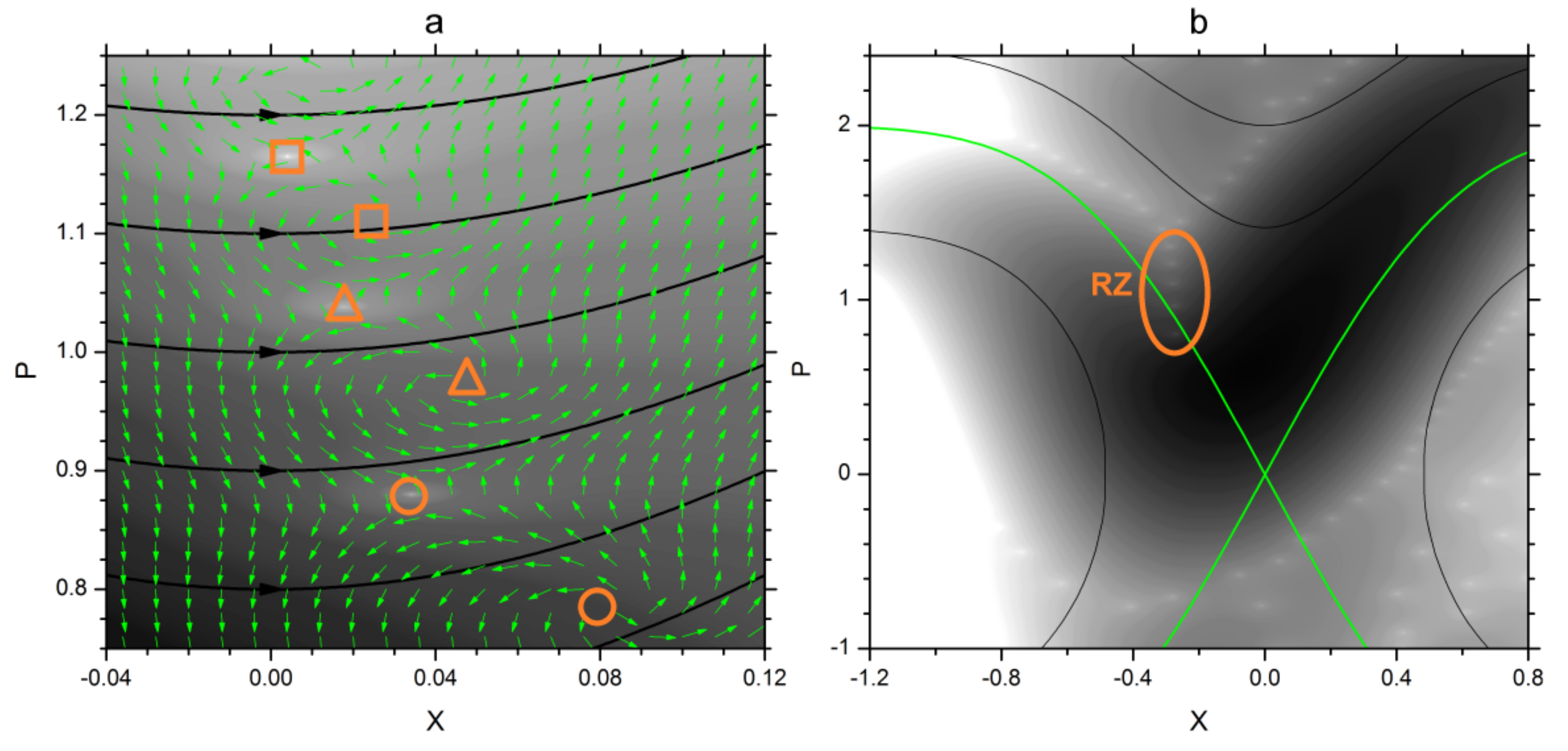}
\par\end{centering}
\caption{{\small (a) Zoom in the row of zeros for $T = 2.1$ and $p_{0} = 1.8$. The black continous lines are the energy levels of the classical Hamiltonian, which coincides with the direction of the classical flow. Three topological dipoles are visible in the image, marked with square, triangle and circle. (b) Husimi function for $p_{0} = 2.1$ and $T = 2.2$. The black continuous lines identify the classical energy levels and the green curve is the separatrix. The row of zeros can be seen crossing the searatrix.}}
\label{fig04}
\end{figure}

In summary, the position of the zeros relative to the Husimi function's maximum and to the classical flow lines are a signature of the transmission regime. This particular model exhibited two particular possibilities of relative position, other systems could offer different situations to be explored.

\section{Final Remarks}

In this work we studied the dynamics of the Husimi function and the role of its zeros in producing stagnation points of the corresponding phase space flow. We showed that the zeros are stagnation saddle points of the quantum probability flow, leading to new topological structures in the current when compared to the classical dynamics. These new stagnation points of the current are created in pairs due to topological restrictions, with indexes equal to $+1$ and $-1$, so that each pair behaves as a topological dipole with total index equal to zero.

As an example we studied the scattering of a wave packet through a Gaussian barrier, where two regimes of transmission can be identified according to the initial average momentum of the particle. For initial average kinetic energy below the barrier top the classical transmission is greater than the quantum one, whereas the quantum transmission is larger than the classical if the average kinetic energy is above the barrier top. In each case the relative position of the topological dipoles is different, accounting for the dynamical mechanism behind the differences between the classical and quantum flow. When the classical transmission is greater than the quantum, the dipoles are located in regions of the phase space where classical trajectories cross the barrier, partially blocking the transmission. When the quantum transmission is greater than the classical, the dipoles are situated near the classical separatrix and they offer a path for probability transmission across it, a classically forbidden mechanism.

The zeros of the Husimi function have already been pointed out as signatures of quantum phenomena. Understanding how these zeros change the phase space probability flow adds new information about the dynamical mechanisms of quantum phenomena in the phase space.

\section*{Acknowledgements}

This work was partly supported by FAPESP and CNPq.

\section*{Appendix}

\paragraph*{Quantum dynamics of the Husimi function}

Given an initial coherent quantum state $\vert z_{0} \rangle$, with mean position and momentum $x_{0}$ and $p_{0}$, respectively, its time evolution is obtained by the split-time operator method (STOM) \cite{bandrauk93}. The output of this method is the wavefunction $\psi \left( x \right)$ in the coordinate representation, and the Husimi function (\ref{eq:definitionQ}) is obtained through a convolution with the coherent state function $\phi^{*} \left( \bar{z} , z ; x \right) = \langle z \vert x \rangle$:
\begin{equation*}
Q \left( \bar{z} , z \right) = \left|\int \psi \left( x \right) \phi^{*} \left( \bar{z} , z ; x \right) \text{d} x \right|^{2}.
\end{equation*}
The quantum current is then evaluated with (\ref{eq:definitionJ2}):
\begin{equation*}
J = \frac{1}{i \hbar} \sum_{l = 1}^{\infty} \frac{\partial^{l - 1} Q}{\partial z^{l - 1}} \sum_{k = 0}^{\infty} \frac{\left(-1 \right)^{k}}{\left(k + l \right)!} \frac{\partial^{2k + l} H}{\partial \bar{z}^{k + l} \partial z^{k}}.
\end{equation*}
In our model the average Hamiltonian (\ref{eq:Haverage}) is not a finite polynomial and for computational purpose we need to truncate the sums. If we consider an $\hbar$ expansion of the current, the contribution of order $\hbar^{N}$ is obtained when $l + k = N + 1$. For example, the classical current has order $\hbar^{0}$, and we would use only the $l = 1$ and $k = 0$ term; the first quantum correction, of order $\hbar^{1}$, would use $l = 2$, $k = 0$ \emph{and} $l = 1$, $k = 1$, and so on. In our calculations we went up to $N=10$.

The STOM routine for the wavefunction was made in the range $-10.0 \leq x \leq 10.0$ with resolution $\Delta x = 0.0025$. The time step was $\Delta t = 0.01$. In Figures 3 and 4 the grid of the Husimi function has resolution of $500 \times 500$ points. In the sequence of images in Figure 3 the region where the Husimi function assumes the greatest values is surrounded by a border of aligned zeros, with an outlying ``sea of zeros''. These zeros were robust-tested and only the border was observed to be accurately reproduced; the position of the zeros in the sea is very sensitive to the range and the resolution of the STOM.

\paragraph*{Classical dynamics of the Husimi function}

The classical evolution of the initial Husimi function was made integrating the trajectories generated by the classical Hamiltonian (\ref{eq:Hclassical}). For time instant $t$, the classical function is
\begin{equation*}
Q \left( x \left( t \right) , p \left( t \right) ; t \right) = Q \left( x , p ; 0 \right),
\end{equation*}
where $\left( x \left( t \right) , p \left( t \right) \right)$ is the trajectory with initial condition $\left( x \left( 0 \right) , p \left( 0 \right) \right) = \left( x , p \right)$, driven by Hamilton's equations of movement:
\begin{equation*}
\dot{x} = \frac{\partial H_{cl}}{\partial p} , \qquad \dot{p} = - \frac{\partial H_{cl}}{\partial x}.
\end{equation*}
In the classical case we used as initial probability density the Husimi function for this state.

\end{document}